\documentclass[aps,prb,twocolumn,showpacs,superscriptaddress]{revtex4}
\usepackage{graphicx,amssymb,amsmath}% Include figure files
\usepackage{natbib}
\usepackage{bm}
\newcommand\bS{\mbox{\boldmath $S$}}
\newcommand\bT{\mbox{\boldmath $T$}}
\newcommand\bq{\mbox{\boldmath $q$}}
\newcommand\CuGeO{CuGeO$_3$~}
\newcommand\CuFeGeO{Cu$_2$Fe$_2$Ge$_4$O$_{13}$}
\newcommand\CuScGeO{Cu$_2$Sc$_2$Ge$_4$O$_{13}$}

\newcommand\RBaNiO{$R_2$BaNiO$_5$}

\newcommand\PbNiVO{PbNi$_2$V$_2$O$_8$}
\begin{document}
\title{Indirect magnetic interaction mediated by spin dimer 
in Cu$_2$Fe$_2$Ge$_4$O$_{13}$}
\author{T. Masuda}
\email[]{tmasuda@yokohama-cu.ac.jp}
\affiliation{International Graduate School of Arts and Sciences,
Yokohama City University, Yokohama, Kanagawa, 236-0027, Japan}

\author{K. Kakurai}
\affiliation{Quantum Beam Science Division, JAEA, Tokai, Ibaraki 319-1195, 
Japan}

\author{M. Matsuda}
\affiliation{Quantum Beam Science Division, JAEA, Tokai, Ibaraki 319-1195, 
Japan}

\author{K. Kaneko}
\affiliation{Advanced Science Research Center, JAEA, Tokai, Ibaraki 319-1195, 
Japan}

\author{N. Metoki}
\affiliation{Advanced Science Research Center, JAEA, Tokai, Ibaraki 319-1195, 
Japan}

\date{\today}

\begin{abstract}
\CuFeGeO\ is a bicomponent compound that consists of 
Cu dimers and Fe chains with separate energy scale. 
By inelastic neutron scattering technique with high-energy resolution 
we observed the indirect Fe - Fe exchange coupling by way of the 
Cu dimers. 
The obtained parameters of the effective indirect interaction and related superexchange interactions 
are consistent with those estimated semi-statically. 
The consistency reveals that the Cu dimers play the role of nonmagnetic media in the 
indirect magnetic interaction. 
\end{abstract}

%\keywords{}

\maketitle

%%%%%%%%%%%%%%%%%%%%%%%%%%%%%%%%%%%%%%%%%%%%%%%%%%
%%%%%%%%%%Text starts here %%%%%%%%%%%%%%%%%%%%%%%
%%%%%%%%%%%%%%%%%%%%%%%%%%%%%%%%%%%%%%%%%%%%%%%%%%
%\section{introduction}
Some types of magnetic interaction in solid are {\it indirect} 
ones by way of nonmagnetic media. 
The superexchange interaction among the most common type 
in insulating metal oxides 
is derived by perturbative treatment of 
electron transfer between nonmagnetic 
anions and magnetic ions~\cite{Anderson59}. 
Another type is found in RKKY interaction 
in magnetic alloys~\cite{RKKY} 
where free electrons in the nonmagnetic metal transfer 
exchange integral between magnetic atoms. 
Anions or nonmagnetic metal behaves as the media for the 
indirect interaction. 
As the exotic types spin-singlet entities such as spin dimers, ladders, 
spin integer chains~\cite{Haldane83} can play similar role. 
The theoretical treatment was considered in 
nonmagnetic impurity doped $S$ = 1 chains~\cite{AKLT87,Miyashita93}, 
$S=1/2$ ladders~\cite{Fukuyama96,Sigrist96}, or 
$S=1/2$ dimerized chains~\cite{Fukuyama96b,Yasuda01}. 
Around the impurity the induced staggered spins generate an effective spin 
of $S=1/2$ in the sea of nonmagnetic state. 
The low energy excitation are described by 
the effective spins that indirectly interact 
with the similar formula to the RKKY interaction. 
This scenario is experimentally realized in doped spin gap compounds 
such as \CuGeO ~\cite{Hase93}, SrCu$_2$O$_3$~\cite{Azuma94}, 
and \PbNiVO ~\cite{Uchiyama99}. 
However, randomness of the impurity distribution makes the system 
rather complex. 
For the basic understanding 
a bicomponent spin compound with 
uniformly distributed magnetic ions is necessary. 

\begin{figure}
\begin{center}
\includegraphics[width=7.5cm]{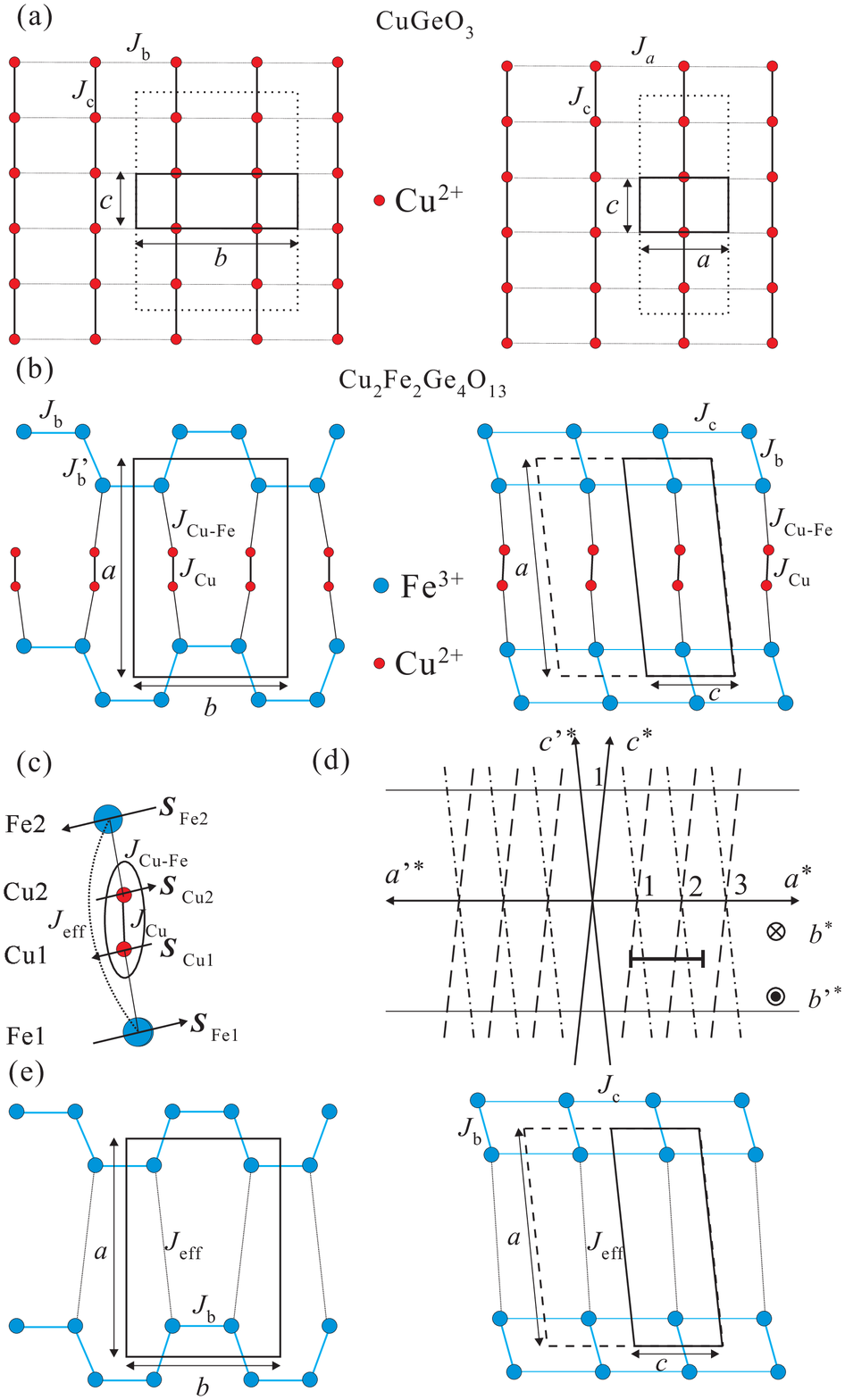}
\end{center}
\caption{(Color online) 
(a) Schematic structure of \CuGeO\ projected onto $c$ - $b$ (left panel) 
and $c$ - $a$ (right panel) planes. 
(b) Schematic structure of \CuFeGeO . 
Lattice unit cell and magnetic unit cell are shown by solid and dashed lines, 
respectively. 
Main exchange path is crank-shaft Fe chains along $b$ direction and 
Cu dimers. 
(c) Indirect Fe - Fe coupling by way of Cu dimers. 
(d) The structure of the twinned crystal of \CuFeGeO\ 
projected onto $a^*$ - $c^*$ plane. 
Dot-dashed and dashed grids are for crystal A and B, respectively. 
(e) Schematic structure of \CuFeGeO\ based on the effective SW model 
(see text). 
} \label{fig1}
\end{figure}

Recently Fe-doped \CuGeO\ with uniform Fe distribution was identified in 
insulating metal oxide \CuFeGeO ; Cu$_{n-2}$Fe$_2$Ge$_n$O$_{3n+1}$ with 
$n=4$ in a hypothetical series of homologous structures with 
$n=3, 5, 6, ... \infty$~\cite{Masuda03a}. 
Here the compound with $n=\infty$ corresponds to well known spin-Peierls cuprite 
\CuGeO ~\cite{Hase93} where uniform 1D chains of Cu$^{2+}$ ions run in 
the $c$ direction as is illustrated in Fig.~\ref{fig1} (a). 
In \CuFeGeO\ 50\% of Cu$^{2+}$ ($S = 1/2$) ions are periodically 
replaced by Fe$^{3+}$ ($S = 5/2$) 
and the uniform chains become distorted alternating chains of 
$S=1/2$ and $S=5/2$ dimers in Fig.~\ref{fig1} (b). 
Consequently the lattice unit is quadrupled along the chain direction and 
the $c$-axis in \CuGeO\ changes to the $a$ axis in \CuFeGeO . 
Bulk susceptibility and neutron inelastic scattering (NIS) measurements indicated that 
the $J_{\rm Cu-Fe}$ are weak~\cite{Masuda04a,Masuda05} 
and, instead, the strength of the interdimer interaction of Fe ions 
($J_{b}'$) is the same as that of the intradimer one ($J_{b}$) within the experimental error. 
Hence \CuFeGeO\ is a weakly coupled quantum dimers of Cu ions and 
classical uniform chains of Fe ions in the $b$ direction. 
In this geometry Cu dimers seem to behave as nonmagnetic media 
for effective exchange interaction ($J_{\rm eff}$) of neighboring Fe ions 
as is described in Fig.~\ref{fig1} (c). 
A set of exchange parameters were determined by NIS and neutron diffraction (ND) 
in the previous study as is summarized in the first line of Table~\ref{parameters}. 
In spite of the rigorous experiments the direct measurement of $J_{\rm eff}$ was not performed 
because of the twinning of the available crystal. 
The values in the square bracket is the reference estimates by a semi-static method. 

In this paper we performed cold neutron inelastic scattering experiment 
with high-energy resolution on the twinned but high-quality crystal. 
Thorough elaborate data analysis we successfully obtained the low energy excitation 
in the $a^{*}$ direction where the $J_{\rm eff}$ makes main contribution. 
We analyze whole dispersion in combination with previously 
reported data~\cite{Masuda05} 
by both the conventional spin wave (SW) and the 
effective SW model based on the energy scale separation. 
We found the latter model more appropriate in quantitative level.
The obtained consistency between the $J_{\rm eff}$'s estimated by 
the dynamics measurement and the semi-static method reveals that the Cu dimers 
play the role of the nonmagnetic media in the indirect interaction between 
Fe spins. 

%\section{Experimental Details}
The high quality crystal with the dimension of 
$7 \times 5 \times 40$ mm$^{3}$ was obtained by 
floating zone method. 
NIS experiment was performed on LTAS spectrometer installed in JRR-3 
in JAEA. 
Open-80'-80'-open with sollar collimation and fixed final energy of 
$E_{f}=3.5$ meV were employed. 
Closed He cycle refrigerator was used to achieve $T = 3.0$ K. 
The crystal structure of \CuFeGeO\ is monoclinic. % and the lattice parameters are 
%$a=12.101$ \AA , $b=8.497$ \AA, $c=4.869$ \AA , and $\beta = 96.131 ^{\circ}$.  
The obtained crystal was twinned so that 
$a^*$ and $b^*$ axes are shared as is illustrated in Fig.~\ref{fig1} (d). 
Here the reciprocal spaces for crystal A and B are defined as $a^* b^* c^*$ 
(dashed grid) and $a^{'*} b^{'*} c^{'*}$ (dashed dotted grid) coordinates, respectively. 
In \CuFeGeO\ a reciprocal vector $(h~k~l)$ in crystal A corresponds to 
$(-(h+0.53l)~-k~l)$ in crystal B. 
The $a^* b^* c^*$ coordinate will be used hereafter. 

\begin{figure}
\begin{center}
\includegraphics[width=7.5cm]{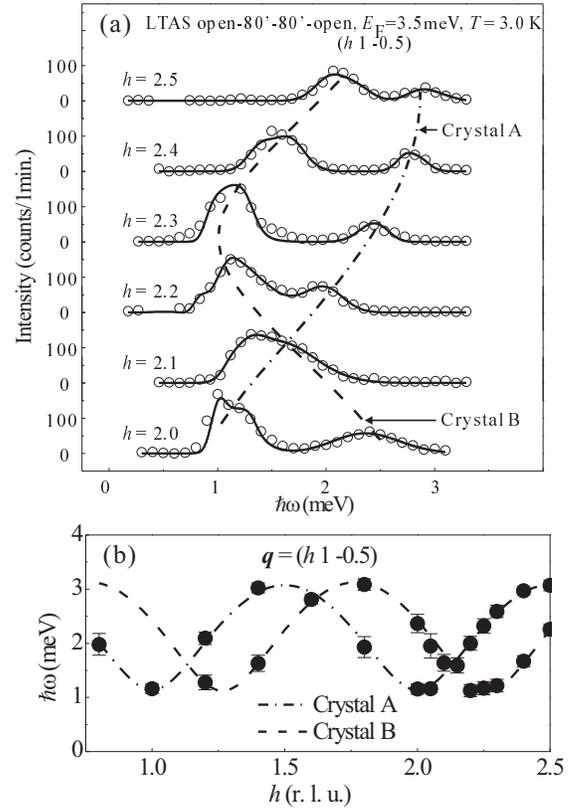}
\end{center}
\caption{
(a) Constant $q$ scans at $(h~1~-0.5)$. 
The dashed-dotted and dashed curves indicate 
excitations from crystal A and B, respectively. 
Solid curves are fitting curves by resolution convoluted 
function. 
(b) Dispersion curves of crystal A (dashed-dot) and 
crystal B (dashed). 
} \label{fig2}
\end{figure}

\begin{figure}
\begin{center}
\includegraphics[width=8.5cm]{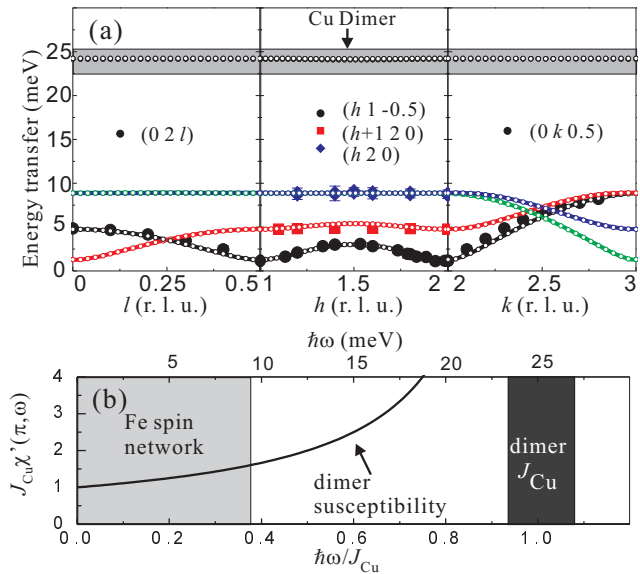}
\end{center}
\caption{(Color online) 
(a) The dispersion relation in \CuFeGeO . 
The lower energy excitation in $a^*$ direction (filled circles) is determined 
in the current study. 
Other data are from Ref.~\onlinecite{Masuda05}. 
Solid lines and small open circles are the fit by conventional SW and effective SW model (see text). 
(b) Energy scales of Fe chains and Cu dimers. 
Dynamic susceptibility of dimers is indicated by a solid curve. 
} \label{fig4}
\end{figure}

%\section{Results}
A series of constant $q$ scans were performed on the 
trajectory of the thick solid line in Fig.~\ref{fig1} (d). 
Peak profiles of the typical scans are shown in Fig.~\ref{fig2} (a). 
At antiferromagnetic zone center of $h$ = 2.0, well-defined spin wave 
excitation with an anisotropy gap of about 1 meV was observed 
in crystal A. 
The peak is wider than experimental resolution and seems to have 
a structure probably due to $xy$ two ion anisotropy. 
The shape of the scan profiles are well explained by 
considering resolution function in the Cooper-Nathans 
approximation~\cite{CooperNathans} that is shown by solid curves. 
At $\hbar \omega \sim 2.4$ meV, broad peak from crystal B 
is detected. 
These two peaks move with 
the change of $h$ on the identical trajectories shown by 
dashed-dotted and dashed curves. 
We fitted the data by a pair of simple Gaussian functions and 
we obtained the dispersion relations in Fig.~\ref{fig2} (b). 
Two curves from crystals A and B are separately 
observed, which are identical by translational 
displacement of $h' = -(h+0.53l)$. 
Thus the dispersion in the $a^*$ direction with the boundary energy of 
about 3 meV is successfully obtained in \CuFeGeO . 
In combination with the previous data~\cite{Masuda05} 
the whole dispersions in three independent directions are drawn in 
Fig.~\ref{fig4} (a). 
Filled symbols are the data by single crystal NIS and 
the gray shaded area around 24 meV is a narrow band excitation 
observed in powder NIS experiment. 

\begin{table*}%[H] add [H] placement to break table across pages
 \caption{Exchange parameters for \CuFeGeO\ determined by neutron inelastic scattering (NIS) 
 and neutron diffraction (ND). The values in the square bracket are the estimates by semi-static method. }
 \label{parameters} 
 \begin{ruledtabular}
 \begin{tabular}{l l l l l l l l}
& $J_{\rm Cu}$~(meV) &  $J_{\rm b}$~(meV) & $J_{\rm c}$~(meV) & $J_{\rm eff}$~(meV) & $J_{\rm Cu-Fe}/J_{\rm Cu}$ 
& $J_{\rm Cu-Fe}$~(meV) & $\Delta$~(meV) \\
 \hline
NIS and ND (Refs.~\onlinecite{Masuda04a} and ~\onlinecite{Masuda05}) & $2.4(2) \times 10$ & 
1.60(2) & 0.12(1) & [0.13(4)] & [0.10(5)] & [2.4(2)] & \\
NIS (Conventional SW) & $2.(4)\times 10^2$ & 1.59(2) & 0.13(2) &  &  & 0.9(6) & 1.0(5) \\
NIS (Effective model) &  & 1.59(2) & 0.13(2) &  0.09(6)  &  & 2.0(5) & 1.0(5)
 \end{tabular}
 \end{ruledtabular}
 \end{table*}

%%%%%%%%%%%%%%%%%%%%%%%%%%%%%%%%%%%%

In the 1st scenario we analyze the data by the conventional SW model. 
Since all the spins are confined in almost $a$ - $c$ plane and the canting is 
small~\cite{Masuda04a}, we start from collinear structure. 
We assume isotropic interactions for Fe - Fe, Cu - Fe, and Cu - Cu spins 
for the simplicity. 
Hence the Hamiltonian is defined as 
\begin{eqnarray}
{\cal H} &=& \frac{1}{2}\sum_{\alpha, \beta, \gamma \in {\rm 8~Fe~ sublattices}, \atop 
\delta, \epsilon, \zeta \in {\rm 8~Cu~ sublattices}} 
\bigl( 
J_{\alpha \beta} \bS _{\alpha}^{\rm Fe} \cdot \bS _{\beta}^{\rm Fe} + 
J_{\gamma \delta} \bS _{\gamma}^{\rm Fe} \cdot \bT _{\delta}^{\rm Cu}  \nonumber \\ 
&+& J_{\epsilon \zeta} \bT _{\epsilon}^{\rm Cu} \cdot \bT _{\zeta}^{\rm Cu}
\bigr).
\label{SWHamiltonian} 
\end{eqnarray}
Here $\alpha$, $\beta$, and $\gamma$ denote Fe sublattices and $\delta$, $\epsilon$, and $\zeta$ 
denote Cu sublattices. 
$\bS^{\rm Fe}$ and $\bT^{\rm Cu}$ are the spin operators of $S=5/2$ and $S=1/2$ on Fe and Cu ions, 
respectively. 
We considered $J_b$, $J_c$, $J_{\rm Cu-Fe}$, and $J_{\rm Cu}$ in Fig.~\ref{fig1} (b) 
as the contributed exchange parameters 
in eq.~(\ref{SWHamiltonian}). 
In addition the split anisotropy gaps $\Delta_x$ and $\Delta_y$ are empirically introduced as 
$\bigl( \hbar \omega_{x,y} (\bq )\bigr) ^2 = \bigl( \hbar \omega (\bq)\bigr)^2+\Delta_{x,y} ^2$. 
$\Delta_{x,y}=$0.95 and 1.16 meV were obtained by fitting the resolution convoluted function to the constant $q$ 
scan at $\bq =$ (2 1 -0.5), where the $xy$ anisotropy is the most enhanced. 
The split is small enough and we will use the mean value $\Delta=1.0(5)$ meV as fixed parameter hereafter. 

The numerical fit to the dispersion is indicated by small open circles in Fig.~\ref{fig4} (a). 
The fit to the data is very good. 
While doing simulation it is found that dispersionless excitation 
at $\hbar \omega \sim 24$ meV, where four modes are overlapped, 
is sensitive to the change of parameter $J_{\rm Cu}$. 
This means that they are from Cu centered spins. 
Similarly four modes in the lower energy are from Fe centered spins. 
%In total eight modes from one magnon emission process are displayed. 
The estimated exchange parameters relating to Fe spins are summarized in 
the second line in Table~\ref{parameters}. 
They are reasonably consistent with previously estimated values 
in the first line. 
However, $J_{\rm Cu}$ of $2.(6) \times 10^2$ meV is very different from the previous value 
24.(2) meV in Ref.~\onlinecite{Masuda05}. 
This is because of the different assumption for the ground state; 
classical N\'eel state for the present SW and singlet dimer state for the previous analysis. 
%This is because SW theory assumes the classical N\'eel state as the ground state. 
%Since the moment on Cu at base temperature is as small as 
%$m_{\rm Cu} = 0.38(4) \mu_{\rm B}$ due to strong qunatum fluctuation~\cite{Masuda04a}, 
%both assumption is not exact. 
%This means that the real ground state is between N\'eel state and 
%dimers singlet state. 
To discuss the $J_{\rm Cu}$ the isostructural compound 
\CuScGeO \cite{Redhammer04} where Fe ions in Fig.~\ref{fig1} (b) 
are replaced by nonmagnetic Sc ions 
would give good insight. 
In these compounds 
the angle $\angle$ Cu-O-Cu and the bond distance are very close to each other 
and so would the magnitude of superexchange constant be. 
The intradimer interaction in \CuScGeO\ was 
reliably determined as $J_{\rm Sc} = 24.(5) \sim 25.(4)$ meV 
by the isolated dimers model~\cite{Masuda06}. 
%As indicated by the crystal structure the typical dimers behaviors 
%were experimentally observed~\cite{Masuda06} with 
%the intradimer interaction of $J_{\rm Sc} = 24.(5) \sim 25.(4)$ meV. 
%The exchange pathway of $J_{\rm Sc}$ in Sc compound 
%corresponds to that of $J_{\rm Cu}$ in Fe compound. 
%The exchange constant in a dimer system can be 
%exactly obtained without any approximation and 
%the value of the $J_{\rm Sc}$ is very reliable. 
%$J_{\rm Cu}$ in 
%the isostructural \CuFeGeO\ would be close to the $J_{\rm Sc}$ estimated above. 
%The estimate of $J_{\rm Cu} = 2.4(2) \times 10$ meV in 
%Ref.~\onlinecite{Masuda05} is quite close to the obtained $J_{\rm Sc}$ 
%compared with the current estimate by SW. 
The value is close to the previous estimate 
of $J_{\rm Cu}$ $24.(2)$ meV~\cite{Masuda05} rather than the present one $2.(6) \times 10^2$ meV. 
Hence, the singlet dimer ground state is a better assumption and 
we adopt the $J_{\rm Cu} = 24.(2)$ meV as the reasonable value hereafter. 

In the 2nd scenario we will consider the effective SW model 
based on the energy scale separation of the magnetic excitations~\cite{Masuda05}. 
The absence of dispersion in the Cu centered mode
and the suppressed magnetic moment means that 
Cu spins are strongly bound to spin dimers singlet state. 
Onto the Cu1 spin in the dimer the field, $h_{\rm Cu1}$, is applied through $J_{\rm Cu-Fe}$ 
by the Fe1 spins in Fig.~\ref{fig1} (c). 
Then the polarized spins, $S_{\rm Cu} = \chi (\pi,\omega)h_{\rm Cu}$, 
are induced on the both Cu1 and Cu2 ions 
that transfer the exchange integral from one Fe spin to another. 
Here the dynamic susceptibility of the dimer $\chi (\pi,\omega)$ is 
purely real and expressed by $\chi (\pi , \omega)=1/(J_{\rm Cu}-\hbar \omega)$ 
at $T = 0$ K (solid curve in Fig.~\ref{fig4} (b)). 
In the energy region of Fe spin networks (gray shaded area) $\chi (\pi , \omega)$ is 
almost $1/J_{\rm Cu}$ and constant. 
This suggests that static treatment of MF-RPA works well when we consider 
the spin dynamics at $\hbar \omega \lesssim 10$ meV. 
Then the indirect coupling $J_{\rm eff}$ 
can be approximately expressed as 
$J_{\rm eff} (\omega ) = J_{\rm Cu-Fe}^2 \chi(\pi , \omega)/2 \sim J_{\rm Cu-Fe}^2/2J_{\rm Cu}$. 
Thus the effective spin Hamiltonian for the low energy region is 
\begin{equation}
{\cal H} = \frac{1}{2}\sum_{\alpha, \beta \in {\rm 8~Fe~ sublattices}} 
J_{\alpha \beta} \bS _{\alpha}^{\rm Fe} \cdot \bS _{\beta}^{\rm Fe}. 
\label{effective} 
\end{equation}
We considered $J_b$, $J_c$, and $J_{\rm eff}$ as the exchange parameters 
as is illustrated in Fig.~\ref{fig1} (e). 
The calculated dispersion relation is 
\begin{widetext}
\begin{equation}
(\hbar \omega _{\bq})^2=
\left\{
   \begin{array}{rl}
    S^2 & \{ (A^2-D^2)-(B^2+C^2+2BC\cos \pi k) \pm 2D\sqrt{B^2+C^2+2BC\cos \pi k} \} \\
    S^2 & \{ (A^2-D^2)-(B^2+C^2-2BC\cos \pi k) \pm 2D\sqrt{B^2+C^2-2BC\cos \pi k} \}
   \end{array}
\right.
\label{dispersion}
\end{equation}
\end{widetext}
Here 
%\begin{eqnarray}
%A &=& 2J_b + J_{\rm eff} + 2J_c \\
%B &=& \sqrt{J_b^2+J_{\rm eff}^2+2J_b J_{\rm eff} \cos 2\pi h} \\
%C &=& J_b \\
%D &=& 2J_c \cos 2\pi l
%\end{eqnarray}
$A = 2J_b + J_{\rm eff} + 2J_c$, 
$B= \sqrt{J_b^2+J_{\rm eff}^2+2J_b J_{\rm eff} \cos 2\pi h}$, 
$C = J_b$, and 
$D = 2J_c \cos 2\pi l.$ 
Reasonable fit to the data is shown by the thick solid curves in Fig.~\ref{fig4} (a). 
They are perfectly overlapped on the open small circles calculated in the 1st scenario. 
The obtained exchange parameters are summarized in the third line in Table~\ref{parameters}. 

Excellent consistency was found in the independently estimated exchange parameters of Fe-centered spins. 
Particularly interesting is the effective indirect coupling $J_{\rm eff}$'s. 
In the current study we obtained $J_{\rm eff} = 0.09(6)$ meV by the 
direct measurement of the {\it spin dynamics}. 
%On the other hand in the previous study $J_{\rm eff}$ was estimated {\it semi-statically}. 
In the previous study~\cite{Masuda04a} $J_{\rm Cu-Fe}/J_{\rm Cu}$ was estimated 
{\it statically} in MF-RPA level 
from the staggered magnetization curves of Cu dimers obtained by ND. 
Combining the parameter with $J_{\rm Cu} = 24.(2)$ meV in a separate experiment~\cite{Masuda05}, 
$J_{\rm eff} = 0.13(4)$ meV was estimated {\it semi-statically}. 
The reasonable consistency in $J_{\rm eff}$'s means that the static MF-RPA treatment works well and, thus, 
Cu dimers behave as the nonmagnetic media in the indirect Fe-Fe coupling. 
Due to the simplicity of the model, all the related exchange constants, $J_{\rm Cu}$, 
$J_{\rm eff}$, and $J_{\rm Cu-Fe}$ are experimentally obtained. 

There are a few other compounds reported as bicomponent systems 
with spin-singlet entity and additional spin. 
Among them \RBaNiO\ ($R$ = rare earth metals) have been studied the most 
in detail~\cite{Zheludev96a}. 
The compound is described by reasonably decoupled one dimensional Ni$^{2+}$ ($S=1$) 
chains and surrounding rare earth ions. 
The weak coupling induce magnetically ordered state of which the nature is similar to 
that in \CuFeGeO\ \cite{Zheludev9801}. 
However, the rare earth moment behaves as Ising type spin and the indirect interaction 
by way of Haldane chains has not been studied. 
Other examples are Cu$_2$CdB$_2$O$_6$ and Cu$_3$(P$_2$O$_6$OH)$_2$~\cite{Hase05}. 
Spin model of quantum dimers and nearly free spins is 
proposed from the observed magnetization plateaux in the $M$ - $H$ curves. 
For lack of single crystal, however, the detailed spin dynamics is still unknown. 
As far as we know \CuFeGeO\ is the only model compound for the indirect magnetic 
interaction by way of spin singlet media. 

%\section{Conclusion}

In conclusion we succeeded to obtain the spin dispersion in 
the $a^*$ direction in \CuFeGeO\ by high-resolution neutron scattering 
experiment and by the elaborate data analysis. 
The whole dispersion is well explained by the effective SW model. 
The effective Fe - Fe interaction $J_{\rm eff}$ is directly 
estimated and the value is consistent with that obtained semi-statically. 
It is found that \CuFeGeO\ is a prototype compound for randomness-free RKKY-like 
interaction in insulating metal oxides. 

%\begin{acknowledgments}
We wish to acknowledge fruitful discussion with Dr. A. Zheludev 
in the early stage of this study. 
Dr. C. Yasuda is appreciated for useful comments. 
This work was supported by the grant for Strategic 
Research Project \#W17003, \#K17028, and \#K18032 of Yokohama City 
University, Japan. 
%\end{acknowledgments}

%\bibliography{CuFeGeOJAEAver1}

\end{document}